\documentclass[a4paper]{article}

\usepackage{INTERSPEECH2020}
\usepackage{amsfonts}
\usepackage{cite}
\usepackage{siunitx}

\title{Multi-path RNN for hierarchical modeling of long sequential data\\ and its application to speaker stream separation}
\name{Keisuke Kinoshita$^1$, Thilo von Neumann$^{2}$, Marc Delcroix$^1$, \\Tomohiro Nakatani$^1$, Reinhold Haeb-Umbach$^2$}
\address{
  $^1$NTT Corporation, Japan,  \ \ \     $^2$Paderborn University, Germany}

\begin{document}

\maketitle
\begin{abstract}
Recently, the source separation performance was greatly improved 
by time-domain audio source separation based on dual-path recurrent neural network (DPRNN).
DPRNN is a simple but effective model for a long sequential data.
While DPRNN is quite efficient in modeling a sequential data of the length of an utterance, i.e., about 5 to 10 second data,
it is harder to apply it to longer sequences such as whole conversations consisting of multiple utterances.
It is simply because, in such a case, the number of time steps consumed by its internal module called inter-chunk RNN becomes extremely large.
To mitigate this problem, this paper proposes a multi-path RNN (MPRNN), a generalized version of DPRNN, that models the input data in a hierarchical manner.
In the MPRNN framework, the input data is represented at several ($\geq3$) time-resolutions, each of which is modeled by a specific RNN sub-module.
For example, the RNN sub-module that deals with the finest resolution may model temporal relationship only within a phoneme,
while the RNN sub-module handling the most coarse resolution may capture only the relationship between utterances such as speaker information.
We perform experiments using simulated dialogue-like mixtures and show that 
MPRNN has greater model capacity, and it outperforms the current state-of-the-art DPRNN framework
especially in online processing scenarios.
\end{abstract}

\section{Introduction}
Automatic meeting analysis is one of the essential technologies required 
for realizing, e.g. communication agents that can follow and respond to our conversations. 
Source separation is one of the important sub-tasks for the meeting analysis.

A considerable number of source separation techniques have been proposed, based on emerging deep learning technologies,
such as Deep Clustering (DC)~\cite{Hershey_ICASSP16}, and Permutation Invariant Training (PIT)~\cite{Yu2016, Kolbaek2017}.
DC and its related technologies~\cite{DAN,Wavesplit} can be viewed as two-stage algorithms. 
They first encode an input signal into an embedding space based on a pretrained neural network (NN), 
and obtain embedding vector(s) for each time frame \cite{Wavesplit} or time-frequency bin \cite{Hershey_ICASSP16,DAN}. 
Then, to obtain source separation masks or separated signals, 
these embedding vectors are clustered by means of e.g. K-means clustering, given the correct number of speakers. 
Since the clustering step is usually used only in the inference stage, 
there is a mismatch in the processing flow between the test and training stage.
Thus, it is difficult for this type of approaches to optimize the total system for source separation metrics.
In contrast, 
PIT and its related approaches \cite{luo2019conv,DPRNN} are single-stage algorithms, 
which let NNs directly estimate separated signals without the clustering step.
Such approaches make it possible not only to optimize the entire system for source separation,
but also to jointly optimize the source separation module with other tasks,
such as a source counting task \cite{RSAN,onlineRSAN_ICASSP2019,Takahashi_Interspeech2019} and an ASR task \cite{Qian2018_Singlechannelmulti,Chang2019_EndEndMonauralMulti,vonNeumann_ICASSP2020}.
Motivated by such an advantage of the single-stage algorithms,
this paper proposes an extension for that approach. 

The PIT-based approach has greatly advanced recently and achieved the state-of-the-art performance 
by incorporating (1) an idea of time-domain audio source separation network (TasNet) framework \cite{luo2019conv}
and (2) an advanced network architecture called dual-path recurrent neural network (DPRNN) \cite{DPRNN}.
The DPRNN framework utilizes RNNs to model a long sequential input in a very simple way.
It first splits the input sequence into short chunks and interleave two RNNs, an intra-chunk RNN and an inter-chunk RNN, for local and global modeling, respectively. 
In a DPRNN block, the intra-chunk RNN first processes the local chunks independently, 
and then the inter-chunk RNN aggregates the information from all the chunks to perform utterance-level processing.
While this algorithm is found to be more efficient than e.g., dilated temporal convolutional NNs 
in modeling sequential data whose length is about an utterance, i.e., about 5 to 10 second data,
it may pose a problem for modeling even longer sequential data.
In such a case, the number of time steps consumed by the inter-chunk RNN becomes extremely large, 
leading to a performance degradation.

Since the modeling of long sequential data is essential when separating long meeting-like data
(i.e., speaker \textit{stream} separation),
this paper proposes to generalize the DPRNN framework to enable efficient modeling of much longer data.
The proposed network architecture, which we call multi-path RNN (MPRNN), models the input sequential data in a hierarchical manner.
In other words, in the MPRNN framework, the input data is represented at several ($\geq3$) time-resolutions, each of which is then modeled by a specific RNN sub-module.
The RNN sub-module dealing with the finest resolution may model temporal relationship within a phoneme,
while the RNN sub-module handling the most coarse resolution may capture the relationship 
between utterances such as speaker information.


\section{Proposed approach: Multi-path RNN}
The proposed MPRNN is a generalized version of the previously proposed DPRNN \cite{DPRNN}, 
which aims to model long temporal context in a sequential input.
In this section, we will explain the core steps in the proposed MPRNN framework, clarifying its relationship to DPRNN.

Similarly to DPRNN, MPRNN consists of mainly two stages; 
segmentation, and a core processing realized using a MPRNN block composed of multiple RNNs.
The segmentation stage splits a sequential input into chunks and forms a tensor that stores the chunks in a hierarchical manner. 
The tensor is then passed to an MPRNN block. 
The MPRNN block is the core of this framework, and models temporal relationships in the input sequential data, 
from short-term (e.g., phoneme level) through middle-term (e.g., an utterance level) to long-term (e.g., inter-utterance level) relationships.
The MPRNN blocks can be stacked on top of each other to enable deep modeling.
An output from the last MPRNN block is transformed back to a sequence data by performing an inverse operation of the segmentation stage and overlap-add.

\subsection{Hierarchical segmentation}
\label{sec:segmentation}
\begin{figure}[t]
	\small
	\centering
	\includegraphics[width=80mm]{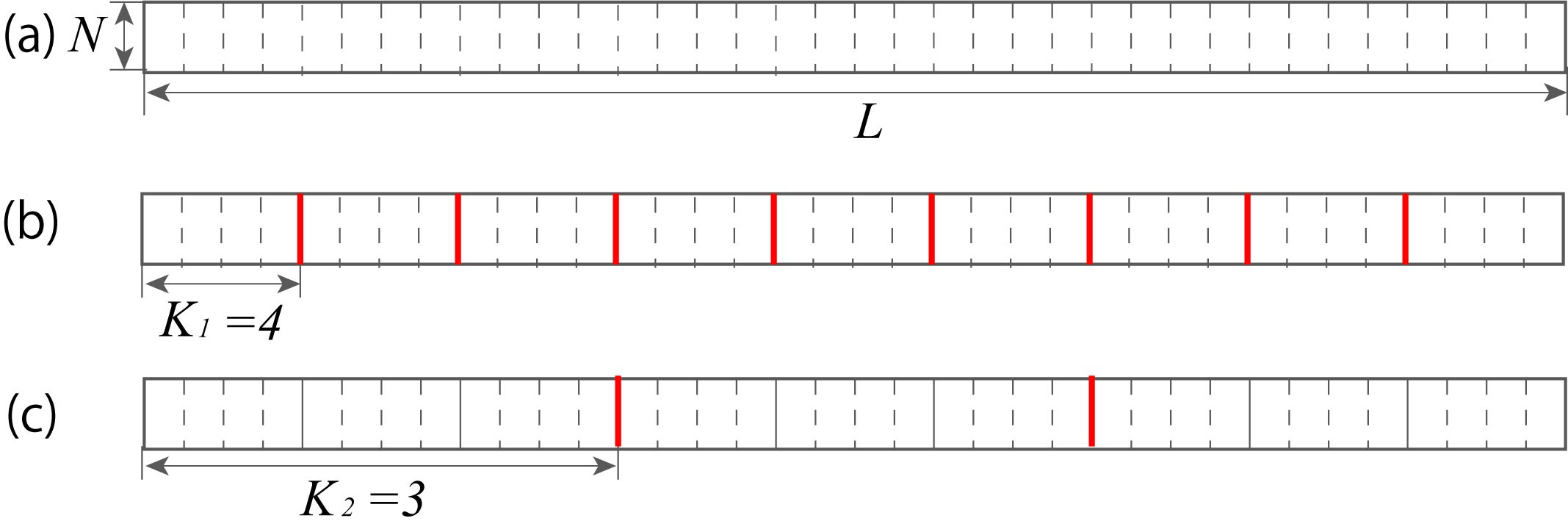}
	\caption{Results of segmentation module in the MPRNN framework: (a) a shape of the input sequence $\vec{W}$ to an MPRNN block. (b) Finest-level segmentation with the chunk length of 4 ($K_1=4$).
	(c) Coarse level segmentation based on (b), with the chunk length of 3 ($K_2=3$). Red lines correspond to boundaries of the chunks.}
	\label{fig:segmentation}
\end{figure}

Let us first denote the input single-channel speech signal in the time domain as $\vec{w} \in \mathbb{R}^{1\times T}$.
Before the segmentation stage, $\vec{w}$ is converted with an encoder convolutional neural network \cite{luo2019conv} to $\vec{W} \in \mathbb{R}^{N\times L}$ 
where $N$ is the feature dimension and $L$ is the number of time frames (see Fig.~\ref{fig:segmentation}~(a)).
Then, the segmentation stage splits $\vec{W}$ into hierarchical chunks as depicted in Fig.~\ref{fig:segmentation}~(b) and (c).
Fig.~\ref{fig:segmentation}~(b) shows the finest-level segmentation applied to the input sequence $\vec{W}$. 
Denoting the length of each chunk in this finest-level segmentation as $K_1$,
we obtain $S_1=L/K_1$ chunks on this segmentation level\footnote{We actually divided the input data into overlapping chunks as in the original DPRNN \cite{DPRNN}, 
but here for the sake of simplicity, we explain a processing flow of MPRNN by assuming the non-overlapping chunks (i.e., corresponding to $P=K$ case in \cite{DPRNN}).}.
Similarly, following the finest segmentation, we apply another segmentation, i.e., more coarse level segmentation on top of the finest segmentation as in Fig.~\ref{fig:segmentation}~(c).
On this segmentation level, a chunk contains several finest-level chunks.
The length of each chunk on this segmentation level is $K_2$ indicating the number of the finest chunks included in a chunk.
On this segmentation level, we obtain $S_2=L/(K_1 \cdot K_2)$ chunks.

Based on this hierarchical segmentation, we can finally form a tensor $T \in \mathbb{R}^{N\times K_1\times K_2\times S_2}$.
Conceptually, by repeating the above hierarchical segmentation $M$ times, 
we can obtain an $(M+2)$-dimensional tensor $T \in \mathbb{R}^{N\times K_1\times K_2\times \dots \times K_M \times  S_M}$.
For the sake of simplicity, we assume $M=2$ case, i.e. $T \in \mathbb{R}^{N\times K_1\times K_2\times S_2}$, for the following explanation.

\begin{figure}[t]
	\small
	\centering
	\includegraphics[width=80mm]{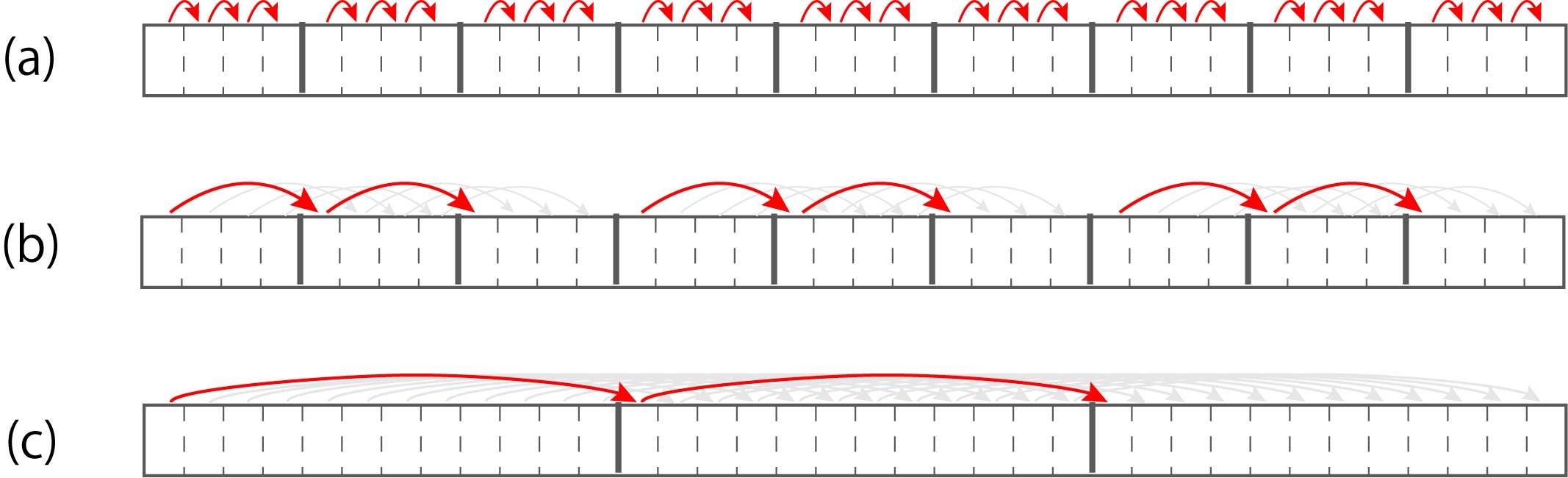}
	\caption{Behavior of RNN sub-modules in an MPRNN block: (a) RNN modeling the finest temporal relationship, (b) RNN modeling the medium-level temporal relationship, (c) RNN modeling the coarse temporal relationship. For the sake of simplicity, uni-directional processing is used for this figure.}
	\label{fig:mprnn_processing}
\end{figure}

\subsection{Processing in an MPRNN block}
The output from the segmentation stage, $\vec{T}$, is then passed to an MPRNN block.
An MPRNN block contains $M+1$ sub-modules each of which models specific temporal relationships in the data by using a (bidirectional) RNN.
In order to process the four-dimensional tensor $T \in \mathbb{R}^{N\times K_1\times K_2\times S_2}$, 
we need 3 sub-modules explained below.
The first, second and third RNN sub-modules aim to roughly capture temporal relationships
within a phoneme, those within an utterance (e.g., inter-phoneme relationships), 
and those between utterances (e.g., speaker information), respectively.

For the RNN sub-module that models the finest temporal relationship, 
the input tensor $\vec{T}$ is reshaped to a three-dimensional tensor $\bar{\vec{T}}_1 \in \mathbb{R}^{N\times K_1 \times (K_2 \cdot S_2)}$.
In $\bar{\vec{T}}_m$ ($m=1,2,3$), the first, second and third dimensions correspond to the input feature size, the number of time steps, and number of samples/examples, in terms of RNN \cite{Tensorflow}.
Using Python-like expressions and notations for matrix operation as in \cite{DPRNN}, 
the finest-level RNN processes the input data $\bar{\vec{T}}_1$ as:
\begin{align}
	\bar{\vec{U}}_1 = [f_1(\bar{\vec{T}}_1[:,:,i]), \, i=1, \ldots, K_2 \cdot S_2], \label{eq:1stRNN}
\end{align}
where $\bar{\vec{U}}_1 \in \mathbb{R}^{H\times K_1 \times (K_2 \cdot S_2)}$ is the output of the RNN, $H$ is the dimension of the hidden layer in the RNN,
$f_1(\cdot)$ is the mapping function defined by an RNN, 
and $\bar{\vec{T}}_1[:,:,i] \in \mathbb{R}^{N\times K_1}$ is the sequence defined by the index $i$.
In other words, the module $f_1(\cdot)$ processes all $K_2 \cdot S_2$ examples independently 
in parallel. 
Figure~\ref{fig:mprnn_processing}~(a) shows how $f_1(\cdot)$ processes the data according to Eq.~(\ref{eq:1stRNN}).
A linear fully-connected (FC) layer is then applied to transform the feature dimension of $\bar{\vec{U}}_1$ back to that of $\bar{\vec{T}}_1$
\begin{align}
	\hat{\vec{U}}_1 = [\vec{G}_1\bar{\vec{U}}_1[:,:,i] + \vec{m}_1, \, i=1, \ldots, K_2 \cdot S_2]
\end{align}
where $\hat{\vec{U}}_1 \in \mathbb{R}^{N\times K_1 \times (K_2 \cdot S_2)}$ is the transformed feature, $\vec{G}_1 \in \mathbb{R}^{N\times H}$ and $\vec{m}_1 \in \mathbb{R}^{N\times 1}$ are the weight and bias of the FC layer.
Finally, the output from this sub-module is formed with a residual connection and layer normalization as:
\begin{align}
	\hat{\vec{T}}_2 = \bar{\vec{T}}_1 + LN(\bar{\vec{U}}_1),\label{eq:residual_connection}
\end{align}
where $LN(\cdot)$ represents a layer normalization \cite{DPRNN}.
$\hat{\vec{T}}_2$ will be used as an input to the following RNN sub-module.
From the explanation of the remaining RNN sub-modules, we omit the processing concerning the FC layer, layer normalization and the residual connection, as it is similar to Eq.~(\ref{eq:residual_connection}). 

For the RNN sub-module that models a medium-level temporal relationship (see Fig.~\ref{fig:mprnn_processing}~(b)),
the tensor obtained in the previous sub-module, $\hat{\vec{T}}_2 \in \mathbb{R}^{N\times K_1 \times (K_2 \cdot S_2)}$, 
is reshaped to $\bar{\vec{T}}_2 \in \mathbb{R}^{N\times K_2 \times (K_1 \cdot S_2)}$.
Then, it is processed as:
\begin{align}
	\bar{\vec{U}}_2 = [f_2(\bar{\vec{T}}_2[:,:,i]), \, i=1, \ldots, K_1 \cdot S_2], \label{eq:2ndRNN}
\end{align}

Similarly, for the RNN sub-module that models a long-term temporal relationship (see Fig.~\ref{fig:mprnn_processing}~(c)),
the tensor obtained from the previous RNN sub-module is reshaped 
to $\bar{\vec{T}}_3 \in \mathbb{R}^{N\times S_2 \times (K_1 \cdot K_2)}$.
Then, it is processed as:
\begin{align}
	\bar{\vec{U}}_3 = [f_3(\bar{\vec{T}}_3[:,:,i]), \, i=1, \ldots, K_1 \cdot K_2], \label{eq:3rdRNN}
\end{align}

\begin{figure}[t]
	\small
	\centering
	\includegraphics[width=50mm]{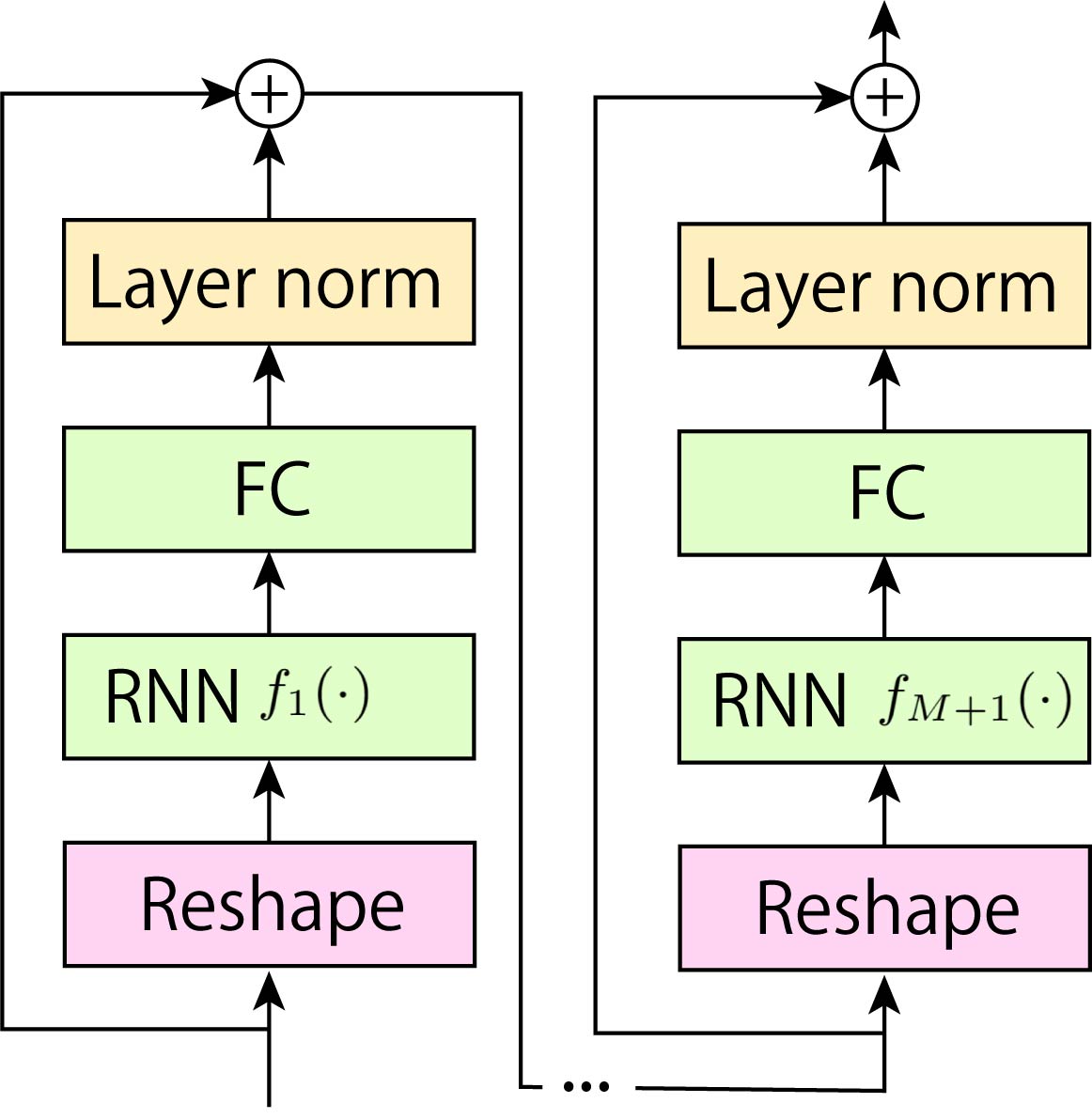}
	\caption{Processing in an MPRNN block. This block can be stacked to enable deep modeling.}
	\label{fig:mprnn_diagram}
\end{figure}

Figure~\ref{fig:mprnn_diagram} shows an overall processing diagram in an MPRNN block containing $M+1$ RNN sub-modules.

\subsection{The relation to DPRNN}
\label{sec:relation}
While MPRNN performs the hierarchical segmentation as explained in section~\ref{sec:segmentation},
segmentation in DPRNN is not hierarchical. It performs the segmentation only one time to obtain the segmentation result depicted in Fig.~\ref{fig:segmentation}~(b).
Thus the obtained tensor in DPRNN is the size of $\mathbb{R}^{N\times K_1\times S_1}$, which is the special case of MPRNN with $M=1$.
Accordingly, DPRNN requires only 2 RNN sub-modules in a DPRNN block which is referred to as inter-chunk and intra-chunk RNNs \cite{DPRNN}. 
On the other hand, MPRNN utilizes M+1 sub-modules.

When modeling a long mixture such as meeting-like data, the number of time steps consumed by the inter-chunk RNN in DPRNN is going to be extremely large,
degrading its performance because LSTM is in general not capable of handling a very long sequence \cite{DPRNN,luo2019conv}. 
On the other hand, MPRNN has extra RNN sub-modules such as $f_3(\cdot)$ that model only long-term temporal relationship in the data (Fig.\ref{fig:mprnn_processing}~(c)).
By skipping the large number of time samples for modeling such a long-term temporal relationship,
MPRNN explicitly circumvent the modeling of extremely large number of time-steps. 
Therefore, we argue that MPRNN is more appropriate to model very long sequential data containing multiple utterances.

\section{Separation framework}
To construct separation networks based on MPRNN and DPRNN, 
we insert the DPRNN and MPRNN modules into the TasNet framework \cite{luo2019conv}.
Specifically, to better understand the behavior of DPRNN and MPRNN, 
we prepared the following two types of frameworks based on TasNet.

\subsection{2-output framework}
The first framework, hereafter referred to as 2-output framework, provides the same outputs as \cite{luo2019conv}, i.e. the models estimate  speech signals for all speakers at once.
We simply replaced the convolutional separation module in \cite{luo2019conv} with the DPRNN or MPRNN modules, keeping the structure of the encoder and decoder unchanged.

\subsection{1-output framework}
Recently, iterative approaches to source separation that output one speaker at a time~\cite{RSAN,onlineRSAN_ICASSP2019,Takahashi_Interspeech2019} have received increased interest as they allow to perform separation for an unknown number of speakers or track speakers along long recordings~\cite{onlineRSAN_ICASSP2019}. We believe that the proposed MPRNN scheme would be particularly suited for such iterative schemes as it may allow better tracking a speaker over a long recording.

As a preliminary investigation of MPRNN for iterative separation, we investigate a second separation framework, referred to as 1-output framework. Here, the network provides a single output speech signal, which consists of one of the speaker arbitrarily chosen by the network. Then, the separated signal for the other speaker is obtained by subtracting the output signal from the mixture in the time domain. Although we limit our investigation to 2-speaker cases in the following experiments, such a scheme could be easily extended to iterative source separation schemes to cope with more speakers \cite{RSAN,Takahashi_Interspeech2019}.


\section{Experimental procedures}
We now evaluate MPRNN in comparison to DPRNN.

\subsection{Data}
For experiments, we generated data based on WSJ0 \cite{WSJ0}.
The training data set for this experiment comprises 20000 examples.
Each example was generated to simulate 30 second dialogue, 
such that every 5 second frame contains zero speaker with a probability of 25~\%,
one speaker with a probability of 50~\%, and two speakers with a probability of 25~\%, respectively.
Development data set comprises 1000 examples, and has similar characteristics to the training data.
For evaluation, we generated two types of evaluation data that differ in length.
The first evaluation data set contains 3000 examples each of which is 30 second long, 
while the second evaluation data set consists of 3000 examples each of which is 120 second long.
Both evaluation data sets were generated by following the same speaker overlap probability used for the training data.

The evaluation metric used in the experiments are SDR \cite{BSSeval}.
The sampling frequency was \SI{8}{\kilo\hertz}.

\subsection{Details of tested DPRNN and MPRNN}
\subsubsection{Model configuration}
For DPRNN, we used the following hyper-parameters. Window size was 16.
Segment length K and hop size were set at 100 and 50, respectively.
The input feature size $N$ and the dimension of the hidden layer $H$ were 64 and 128.
The number of the DPRNN blocks used for this experiment was 5.
With these settings, the number of time-steps consumed by the inter-chunk RNNs in DPRNN are
600 for 30 second data, and 2400 for 120 second data, respectively.

For MPRNN, we evaluated the case where the number of hierarchical segmentation was two, i.e. $M=2$.
Segment length $K_1$ and $K_2$ were set at 100 and 60, and corresponding hop sizes were set at 50 and 30, respectively.
Other parameters were kept the same as the DPRNN setting for a fair comparison. The number of MPRNN blocks was set to 3 to have a total number of parameters similar to the baseline DPRNN model containing 5 DPRNN blocks.

We train all models for 150 epochs with utterance-level permutation invariant training \cite{Kolbaek2017} to maximize scale-dependent signal-to-distortion ratio (SD-SDR). 
The network was trained on a whole utterance of 30 seconds, rather than a part of an utterance.
Adam is used as the optimizer, with an initial learning rate of 0.001. 
For all configurations, we used the model that achieved the best performance on the validation set. 




\subsubsection{Latency}
\label{sec:latency}
When processing relatively long data, online processing is often needed.
To clarify how the algorithm latency affects the performance,
we prepared offline and online settings for DPRNN and MPRNN models.

In case of the offline DPRNN, we used BLSTM networks for both inter-chunk and intra-chunk RNNs.
For the online DPRNN, we used BLSTM networks for intra-chunk RNNs, and LSTM networks for inter-chunk RNNs.

Similarly, in case of the offline MPRNN, we use BLSTM networks for all RNNs from $f_1(\cdot)$ to $f_3(\cdot)$.
For online MPRNN, we use BLSTM for $f_1(\cdot)$ and $f_2(\cdot)$, and LSTM for $f_3(\cdot)$.
By doing so, the online DPRNN and MPRNN models have the same number of parameters.

With the aforementioned parameters, 
the online MPRNN model has an algorithmic delay of about 1.5 seconds, 
while the online DPRNN model works with the delay of 100~ms.
To make a fair comparison with online MPRNN, we investigated increasing the delay for DPRNN by delaying the output,
but we found that DPRNN would not train well in such cases.
Therefore, in our investigation, we employ the online DPRNN model with the delay of 100~ms as the best performing online DPRNN model.

\section{Results}

\begin{table}[t]
\centering
 \caption{Performance of offline DPRNN and MPRNN for 30 and 120 second data, in the 1-output
          and 2-output frameworks}
          \vspace{-1mm}
  \label{tbl:results_offline_longdata}
  \begin{tabular}{l  l  c c}
    \toprule
    \multicolumn{2}{c}{Model} &  \multicolumn{2}{c}{SDR (dB)}  \\  
    Separator & Framework   & 30 sec.     & 120 sec.    \\ \hline \hline  \\[-2.3ex]
    DPRNN     & 2-output    &  19.61      & 16.15  \\
              & 1-output    &  19.32      & 15.75  \\ \hline
    MPRNN     & 2-output    &  19.39      & 15.90  \\
              & 1-output    &  \textbf{19.72}      &  \textbf{16.26}           \\ \hline
  \end{tabular}
\end{table}

\begin{figure}[t]
	\small
	\centering
	\includegraphics[width=80mm]{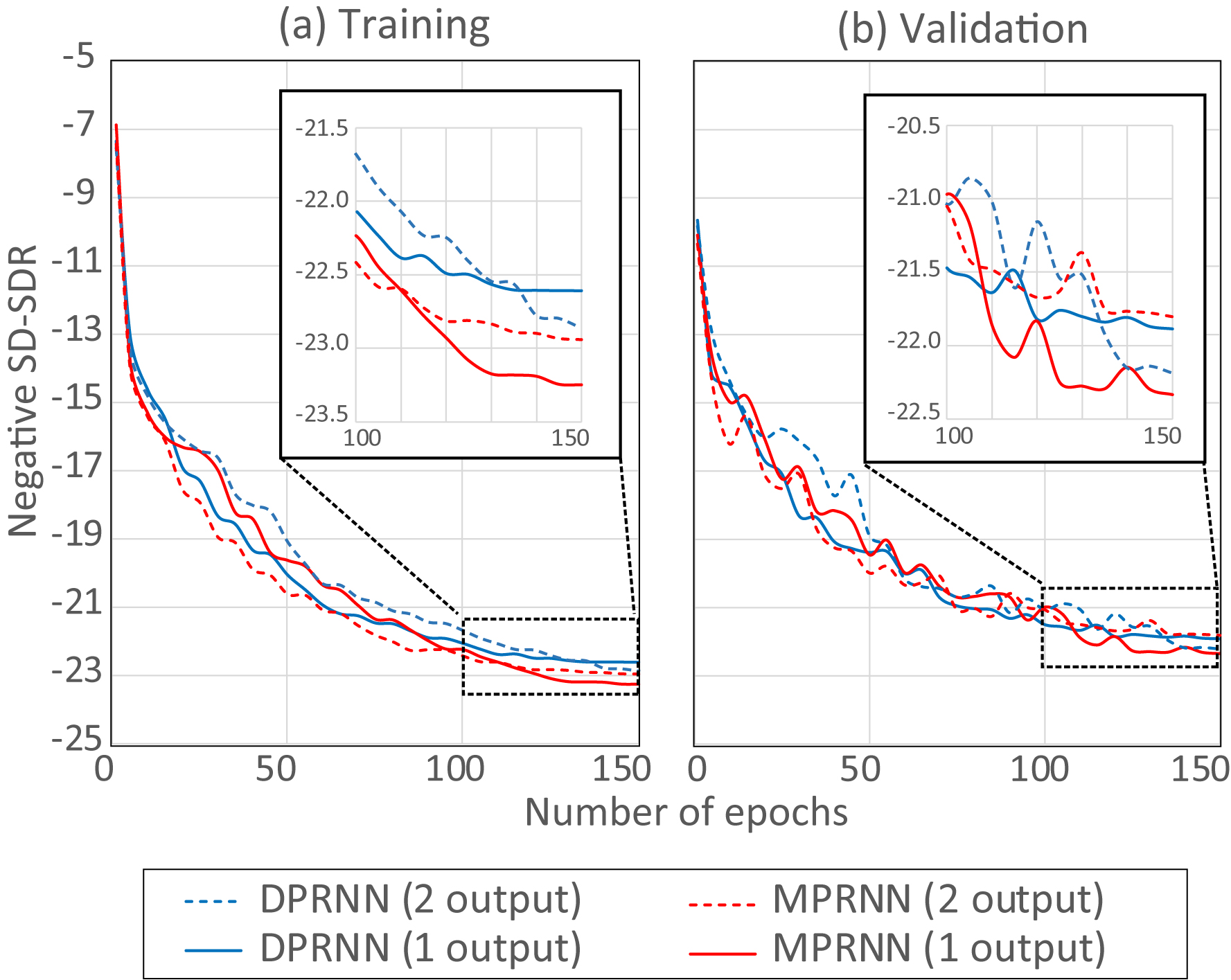}
	\caption{Loss curves for (a) training and (b) validation data.}
	\label{fig:loss_curve}
\end{figure}

\subsection{Offline processing}
Table~\ref{tbl:results_offline_longdata} shows the separation results obtained with offline DPRNN and offline MPRNN in case of 2-output and 1-output frameworks
for 30 and 120 second data sets.
The MPRNN with the 1-output framework slightly outperforms the other models.  

To further analyze the performance and capacity of each model,
we show loss curves for the training and validation data in Fig.~\ref{fig:loss_curve}.
Looking at the loss curves for the training data, 
we found that the MPRNN models tend to converge to lower SD-SDR values, 
which may suggest greater model capacity of MPRNNs.
Note that, the number of parameters for the offline DPRNN and MPRNN models are 2.17M and 1.95M, respectively. 
In other words, the greater model capacity of MPRNN does not come from the size of the model,
but comes from the inherent structure of the model.

However, the loss curves for the validation data do not always coincide with the training loss curves.
Specifically, the superiority of MPRNNs for the training data did not necessarily translate to
the improvement for the validation and evaluation data.
This mismatch may suggest that MPRNNs overfit to the training data.
This overfitting issue will be revisited in our future work, by e.g., increasing the amount of the training data,
and improving the model training scheme.


\subsection{Online processing}
Table~\ref{tbl:results_online_longdata} shows the results obtained with the online DPRNN and online MPRNN models.
In this case, the MPRNN models outperform the DPRNN models,
capitalizing on its RNN sub-modules that model long-term temporal relationships.
However the performance of the DPRNN models should be interpreted carefully 
since the algorithmic delay between the DPRNN and MPRNN models are different (See section~\ref{sec:latency}).
Comparing the performance difference between the offline and online models,
we see that the performance degradation for the online MPRNN models are smaller than with the online DPRNN models,
showing the potential of MPRNN for block-online source separation task \cite{onlineRSAN_ICASSP2019}.

Looking at the results with the 120 second data set,
we observe performance degradation for both DPRNN and MPRNN when the test data is much longer than the training data. 
However, the MPRNN still maintains its superiority and achieves high separation performance of \SI{14.82}{\decibel}.

\begin{table}[t]
\centering
 \caption{Performance of online DPRNN and MPRNN for 30 and 120 second data, in the 1-output
          and 2-output frameworks}
          \vspace{-1mm}
  \label{tbl:results_online_longdata}
  \begin{tabular}{l  l  c c}
    \toprule
    \multicolumn{2}{c}{Model}       &  \multicolumn{2}{c}{SDR (dB)}  \\  
    Separator & Framework   & 30 sec.     & 120 sec.    \\ \hline \hline  \\[-2.3ex]
    DPRNN     & 2-output    &  15.69      & 13.22  \\
              & 1-output    &  16.73      &  14.01  \\ \hline
    MPRNN     & 2-output    &  17.70      & 14.62  \\
              & 1-output   &  \textbf{17.90}      &  \textbf{14.82}          \\ \hline
  \end{tabular}
\end{table}

\section{Conclusions}
This paper proposed an efficient sequence model, called MPRNN, and described its application to the source separation task.
The proposed MPRNN is a generalization of DPRNN that achieves the state-of-the-art performance.
In the MPRNN framework, the input data is represented at several time-resolutions. 
Then, the data in each resolution is modeled by a specific RNN sub-module.
Experimental results suggests (1) MPRNN potentially has greater model capacity, 
and (2) it outperforms DPRNN especially in online processing scenarios.


\bibliographystyle{IEEEtran}

\bibliography{main}


\end{document}